\documentclass[prd,aps,showpacs,nofootinbib]{revtex4}%
\usepackage{graphicx}
\usepackage{amsmath}
\usepackage{amsfonts}
\usepackage{amssymb}%
\newcommand{\be}{\begin{equation}}
\newcommand{\ee}{\end{equation}}
\newcommand{\bq}{\begin{eqnarray}}
\newcommand{\eq}{\end{eqnarray}}
\begin{document}
\title{Influence of Lorentz-violating terms on a two-level system}
\author{Manoel M. Ferreira Jr$^{a,c}$, Adalto R. Gomes$^{b}$, Rafael C. C. Lopes$^{a}$}
\affiliation{$^{a}$Departamento de F\'{\i}sica, Universidade Federal do Maranh\~{a}o
(UFMA), Campus Universit\'{a}rio do Bacanga, S\~{a}o Lu\'{\i}s-MA, 65085-580 - Brazil}
\affiliation{$^{b}$ Departamento de Ci\^{e}ncias Exatas, Centro Federal de Educa\c{c}\~{a}o
Tecnol\'{o}gica do Maranh\~{a}o - CEFET-MA S\~{a}o Lu\'{\i}s-MA, 65025-001, Brazil}
\affiliation{$^{c}${\small {Grupo de F\'{\i}sica Te\'{o}rica Jos\'{e} Leite Lopes, C.P.
91933, CEP 25685-970, Petr\'{o}polis, RJ, Brazil}}}

\begin{abstract}
The influence of \ Lorentz- and CPT-violating terms of the extended Standard
Model on a semi-classical two-level system is analyzed. It is shown that the
Lorentz-violating background (when coupled with the fermion sector in a vector
way) is able to induce modifications on the Rabi oscillation pattern,
promoting sensitive modulations on the usual oscillations. As for the term
involving the coefficient coupled in an axial vector way, it brings about
oscillations both on energy states and on the spin states (implied by the
background). It is also seen that such backgrounds are able to yield state
oscillations even in the absence of the electromagnetic field. The foreseen
effects are used to establish upper bounds on the Lorentz-violating coefficients.

\end{abstract}
\email{manojr@pq.cnpq.br, argomes@pq.cnpq.br}

\maketitle

\section{Introduction}

Planck scale physics is an unknown frontier where gravitational and quantum
effects are closely entwined. At this scale, it might occur that Lorentz
covariance is jeopardized. Such kind of idea has caught much attention mainly
after some authors argued the possibility of \ Lorentz and CPT spontaneous
breaking in the context of string theory \cite{Samuel}. The detection of
Lorentz violation at a lower energy scale, even minuscule, could be
interpreted as a signature of spontaneous Lorentz violation at the underlying
theory (defined at\ a higher energy scale). These\textbf{ }remanent\textbf{
}Lorentz violating effects, inherited from a high energy theory, would be then
employed to indicate possible features of a Planck scale physics.\textbf{ }The
Standard Model Extension (SME) \cite{Colladay} is a broader version of the
usual Standard Model that incorporates all Lorentz-violating
(LV)\ coefficients (generated as vacuum expectation values of the underlying
theory tensor quantities) that yield Lorentz scalars (as tensor contractions)
in the observer frame. Such coefficients govern Lorentz violation in the
particle frame, where are seen as sets of independent numbers.

The SME is actually the suitable framework to investigate properties of
Lorentz violation on physical systems involving photons \cite{photons1},
\cite{photons2}, radiative corrections \cite{Radiative}, fermions
\cite{fermions}, neutrinos \cite{neutrinos}, topological defects
\cite{Defects}, topological phases \cite{Phases}, cosmic rays \cite{CosmicRay}%
, supersymmetry \cite{Susy}, particle decays \cite{Iltan}, and other relevant
aspects \cite{Lehnert1}, \cite{General}. The SME has also been used as
framework to propose Lorentz violating \cite{Tests} and CPT\ \cite{CPT}
probing experiments, which have amounted to the imposition of stringent bounds
on the LV coefficients.

Concerning the fermion sector of the SME, there are two CPT-odd terms,
$v_{\mu}\overline{\psi}\gamma^{\mu}\psi,b_{\mu}\overline{\psi}\gamma_{5}%
\gamma^{\mu}\psi$, where $v_{\mu},b_{\mu}$ are the LV backgrounds. The
influence of these terms on the Dirac theory has already been examined in
literature \cite{Hamilton}, passing through its nonrelativistic limit, with
close attention on the hydrogen spectrum \cite{Manojr}. A similar study has
also been developed for the case of a non-minimal coupling with the
background, with new outcomes \cite{Nonmini}.\textbf{ }Atomic and optical
physics is another area in which Lorentz violation has been intensively
studied. Indeed, there are several works examining Lorentz violation in
electromagnetic cavities and optical systems \cite{Cavity}, \cite{Masers},
which contributed to establish upper bounds on the LV coefficients.

The present work is devoted to investigating the influence of Lorentz
violation induced by the coefficients $\ v_{\mu},b_{\mu}$ on the physics of a
semi-classical two-level system. As some fundamental concepts of two-level
system are also important for the description of laser systems, some results
obtained at a semi-classical level may indicate perspectives on the quantum
behavior of photons on a resonant cavity. We start from the Lorentz-violating
nonrelativistic Hamiltonian stemming from the Dirac Lagrangian supplemented
with the terms $v_{\mu}\overline{\psi}\gamma^{\mu}\psi,b_{\mu}\overline{\psi
}\gamma_{5}\gamma^{\mu}\psi.$ The LV terms are then considered as
perturbations that may modify the dynamics of Rabi oscillations on a two-level
system. \ The first analysis is performed for the term $v_{\mu}\overline{\psi
}\gamma^{\mu}\psi.$ It is seen that it induces modifications on the population
inversion function (PIF), that may be partially frustrated in some situations
or modulated as a beat for other parameter values. Numerical simulations
indicated the absence of LV effects on the system for $v_{x}\leq10^{-10}eV$,
which thus can be taken as an upper bound for this background.\textbf{ }In
order to examine the effect of the term $b_{\mu}\overline{\psi}\gamma
_{5}\gamma^{\mu}\psi,$ we have defined a four-state basis, considering the
possibility of the electron spin to be up or down. As a consequence, both
eigenenergy and spin state oscillations take place.\ These backgrounds are
able to induce Rabi oscillations even in the absence of electromagnetic
external field. The non observation of spin oscillation in a real situation
was used to set up an upper bound on the $\mathbf{b}$ magnitude ($|\mathbf{b}%
_{x}|<10^{-19}eV).$

This paper is outlined as follows.\ In Sec. II, it is presented the fermion
sector Lagrangian here taken into account, with the associated nonrelativistic
Hamiltonian. In Sec. III, some topics of a two-level system are revisited.
Further, the Lorentz-violating effects on such a system are discussed and
analyzed. In Sec. IV, we finish with our concluding remarks.

\section{Lorentz-violating Dirac Lagrangian}

\qquad We begin considering the presence of the two Lorentz- and CPT-violating
terms ($v_{\mu}\overline{\psi}\gamma^{\mu}\psi,b_{\mu}\overline{\psi}%
\gamma_{5}\gamma^{\mu}\psi)$ in the fermion sector,
\begin{equation}
\mathcal{L}%
\acute{}%
\mathcal{=L}_{Dirac}-v_{\mu}\overline{\psi}\gamma^{\mu}\psi-b_{\mu}%
\overline{\psi}\gamma_{5}\gamma^{\mu}\psi, \label{L1}%
\end{equation}
where $\mathcal{L}_{Dirac}$ is the usual Dirac Lagrangian ($\mathcal{L}%
_{Dirac}=\frac{1}{2}i\overline{\psi}\gamma^{\mu}\overleftrightarrow{\partial
}_{\mu}\psi-m_{e}\overline{\psi}\psi)$, $v_{\mu}$ and $b_{\mu}$ are two
CPT-odd coefficients that here represent the fixed background responsible for
the violation of Lorentz symmetry in the frame of particles. In true, the
terms $v_{\mu}\overline{\psi}\gamma^{\mu}\psi$, $b_{\mu}\overline{\psi}%
\gamma_{5}\gamma^{\mu}\psi$ behave as a scalar and a pseudoscalar only in the
observer frame, in which $v_{\mu}$\ and $b_{\mu}$ are seen as genuine
4-vectors and no Lorentz-violation takes place\cite{Colladay}. The
Euler-Lagrange equation applied on this Lagrangian provides the modified Dirac equation:%

\begin{equation}
\left(  i\gamma^{\mu}\partial_{\mu}-v_{\mu}\gamma^{\mu}-b_{\mu}\gamma
_{5}\gamma^{\mu}-m_{e}\right)  \psi=0, \label{DiracS1}%
\end{equation}
which corresponds to the usual Dirac equation supplemented by the
Lorentz-violating terms associated with the background. Such equation is also
attainable in the momenta space:
\begin{equation}
\left(  \gamma^{\mu}p_{\mu}-v_{\mu}\gamma^{\mu}-b_{\mu}\gamma_{5}\gamma^{\mu
}-m_{e}\right)  w(p)=0, \label{DiracS2}%
\end{equation}
with $w\left(  p\right)  $ being the $\left(  4\times1\right)  $ spinor in
momentum space. It is possible to show that each component of the spinor $w$
satisfies a changed Klein-Gordon equation which represents the dispersion
relation of this model, given as follows:%
\begin{equation}
\left[  \left[  \left(  p-v\right)  ^{2}-b^{2}-m^{2}\right]  ^{2}%
+4b^{2}(p-v)^{2}-4[b\cdot(p-v)]\right]  =0, \label{DR1}%
\end{equation}

We now asses the nonrelativistic limit of such modified Dirac equation. To
correctly do it, Lagrangian (\ref{L1}) must be considered in the presence of
an external electromagnetic field $\left(  A_{\mu}\right)  $ coupled to the
matter field by means of the covariant derivative ($D_{\mu}=\partial_{\mu
}+ieA_{\mu}).$ Lagrangian (\ref{L1}) is then rewritten in the form:
\begin{equation}
\mathcal{L=}\frac{1}{2}i\overline{\psi}\gamma^{\mu}\overleftrightarrow{D}%
_{\mu}\psi-m_{e}\overline{\psi}\psi-b_{\mu}\overline{\psi}\gamma_{5}%
\gamma^{\mu}\psi-v_{\mu}\overline{\psi}\gamma^{\mu}\psi, \label{DiracS3}%
\end{equation}
which implies: $\left(  \gamma^{\mu}\partial_{\mu}-e\gamma^{\mu}A_{\mu}%
-v_{\mu}\gamma^{\mu}-b_{\mu}\gamma_{5}\gamma^{\mu}-m_{e}\right)  w(p)=0.$
Writing $w(p)$ in terms of two-component spinors ($w_{A},w_{B}),$ there
follows: \qquad%
\begin{align}
\left(  E-eA_{0}-m_{e}-\mathbf{\sigma}\cdot\mathbf{b}-\text{v}_{0}\right)
w_{A}  &  =\left[  \mathbf{\sigma}\cdot(\mathbf{p}-e\mathbf{A}-\mathbf{v}%
)-b_{0}\right]  w_{B},\label{WA2}\\
\left(  E-eA_{0}+m_{e}-\text{v}_{0}-\mathbf{\sigma}\cdot\mathbf{b}\right)
w_{B}  &  =\left[  \mathbf{\sigma}\cdot(\mathbf{p}-e\mathbf{A}-\mathbf{v}%
)-b_{0}\right]  w_{A}. \label{WB2}%
\end{align}
$\ \ \ \ $\

These two equations yield the free-particle solutions for this model (see
refs. \cite{Colladay},\cite{Manojr}). The non-relativistic regime is realized
by the well-known conditions $\mathbf{p}^{2}\ll m_{e}^{2},$ $eA_{0}\ll
m_{e},E=m_{e}+H,$ where $H$ represents the nonrelativistic Hamiltonian.
$\ $Now, replacing the small spinor component ($w_{B})$ on the eq.
(\ref{WA2}), it is attained an equation for the large spinor component
($w_{A})$ that governs the behavior of the system at this regime. Such
equation also provides an expression for $H$ (see ref. \cite{Manojr}):%
\begin{equation}
H=H_{\text{Pauli}}+\left[  -\frac{(\mathbf{p}-e\mathbf{A})\cdot\mathbf{v}%
}{m_{e}}+\mathbf{\sigma}\cdot\mathbf{b}-\frac{b_{0}}{m_{e}}\mathbf{\sigma
}\cdot(\mathbf{p}-e\mathbf{A})\right]  .
\end{equation}
To properly study the influence of this Hamiltonian on a quantum system, the
Lorentz-violating terms (into brackets) should be considered into the
Schr\"{o}dinger equation. In the next section, it will be accomplished for a
two-level system.

\section{Effects on a two-level system}

\subsection{Typical description of Rabi oscillation on a two-level system}

Consider a two-level system defined by the energy eigenstates $|a\rangle
,|b\rangle,$ under the action of a semiclassical electromagnetic field
$(A^{\mu}).$ The wavefunction for this system is:
\begin{equation}
|\psi(t)\rangle=A(t)|a\rangle+B(t)|b\rangle,
\end{equation}
so that $|A(t)|^{2}$, $|B(t)|^{2}$ represent the probability of finding the
electron in the states $|a\rangle$, $|b\rangle,$ respectively. The evolution
of this system is given by the Schr\"{o}dinger equation,
\begin{equation}
|\overset{\cdot}{\psi}(t)\rangle=-iH|\psi(t)\rangle,\label{Schro}%
\end{equation}
with $H$ \ being the associated Hamiltonian, which may be written in terms of
an unperturbed and an interaction part, namely: $\ H=H_{0}+H_{int}$, where
\ $H_{0}|a\rangle=\hbar\omega_{a}|a\rangle,$ $H_{0}|b\rangle=\hbar\omega
_{b}|b\rangle,$ and $H_{int}=-e\mathbf{r}\cdot\mathbf{E}(\mathbf{r},t),$
whereas\textbf{ }$\mathbf{r}$ concerns to the atom position. For the case when
the electric field is polarized along the x-axis, $\overrightarrow{E}%
(t)=E_{0}\cos\nu t\widehat{i},$ we get the result $H_{int}=-exE_{0}\cos\nu t.$
Observe that the electric field modulus ($E_{0})$ is taken as a constant. This
is a consequence of the dipole approximation (see ref. \cite{Scully}).

In order to know how the electric field acts on the system, we should
determine the state vector $|\psi(t)\rangle,$ that is, the coefficients $A(t)$
and $B(t).$ For this, we write the Hamiltonian in the basis of states
$\{|a\rangle$ , $|b\rangle\}$: $\ H_{0}=\hbar\omega_{a}|a\rangle\langle
a|+\hbar\omega_{b}|b\rangle\langle b|,$ $H_{int}=-(P_{ab}|a\rangle\langle
b|+P_{ab}^{\ast}|b\rangle\langle a|)E(t),$ where $P_{ab}=e\langle
a|x|b\rangle$ is the matrix element of the electric dipole moment. Replacing
the Hamiltonian and the ket $|\psi(t)\rangle$ in the Schr\"{o}dinger equation,
two coupled differential equations for $A(t),B(t)$ arise:%
\begin{align}
\overset{\cdot}{A}(t)  &  =-iA\omega_{a}+i\Omega_{R}B\cos\nu t,\label{ca1}\\
\overset{\cdot}{B}(t)  &  =-iB\omega_{b}+i\Omega_{R}A\cos\nu t, \label{cb1}%
\end{align}
where $\Omega_{R}=|P_{ab}|E_{0}$\ is the Rabi frequency and $P_{ab}$ is here
supposed to be a real function. Equations (\ref{ca1},\ref{cb1}) may be easily
solved for the slowly varying amplitudes, $a(t)=Ae^{i\omega_{a}t}%
,b(t)=Be^{i\omega_{b}t}$, with which they are read as:
\begin{align}
\overset{\cdot}{a}(t)  &  =i\frac{\Omega_{R}}{2}b(t)e^{i(\omega-\nu
)t},\label{a1}\\
\overset{\cdot}{b}(t)  &  =i\frac{\Omega_{R}}{2}a(t)e^{-i(\omega-\nu)t}.
\label{b1}%
\end{align}
Here $\omega=(\omega_{a}-\omega_{b}),$ and we used the rotating wave
approximation (RWA), in which the rapidly oscillating terms, $\exp[\pm
i(\omega+\nu)t],$ were neglected. Under this approximation the equations for
$a(t)$ and $b(t)$ may be exactly solved. Considering the system in the state
$|a\rangle$ at $t=0,$ we get the results
\begin{align}
a(t)  &  =\left[  \cos(\Omega t/2)-i(\Delta/\Omega)\sin(\Omega t/2)\right]
e^{i\Delta t/2},\label{sol1a}\\
b(t)  &  =i\frac{\Omega_{R}}{\Omega}\sin(\Omega t/2)e^{-i\Delta t/2},
\label{sol1b}%
\end{align}
with $\Delta=(\omega-\nu),\Omega=\sqrt{\Omega_{R}^{2}+(\omega-\nu)^{2}}.$ At
resonance, the frequency of the external field coincides with the two-level
frequency difference $\left(  \nu=\omega\right)  ,$ so that $\Delta
=0,\Omega=\Omega_{R}.$ For more details, see ref. \cite{Scully}. The
population inversion function (PIF), defined as $W(t)=|a(t)|^{2}-|b(t)|^{2},$
is then equal to%
\begin{equation}
W(t)=\cos\Omega_{R}t. \label{W2}%
\end{equation}
It varies from $-1$ and $1$, reflecting the alternation of the particle
between the states $|a\rangle$, $|b\rangle$ along the time.

\subsection{Lorentz-violating effects due to the vector coupling}

Our first step is to determine the role played by the Hamiltonian terms
stemming from $v_{\mu}\overline{\psi}\gamma^{\mu}\psi$ on the two-level
system, whose effect is governed by the nonrelativistic terms $e\mathbf{A}%
\cdot\mathbf{v}/m_{e}$ and $\mathbf{p}\cdot\mathbf{v}/m_{e}$. We begin by
regarding the effect of the term $\mathbf{A}\cdot\mathbf{v}$. This can be done
following the procedure of the last section. Taking on $\mathbf{E}%
(t)=E_{0}\cos\nu t\widehat{i},$ it results $\mathbf{A}(t)=-A_{0}\sin\nu
t\widehat{i},$ with $A_{0}=E_{0}/$ $\nu.$ The Hamiltonian reads as
$H=H_{0}+H_{int}+H_{1},$ where $H_{1}=-eA_{0}$v$_{x}\sin\nu t/m_{e}.$ In the
basis $\{|a\rangle$ , $|b\rangle\},$ $H_{1}$ has the simple form:
$H_{1}=\alpha(t)(|a\rangle\langle a|+|b\rangle\langle b|),$ with:
$\alpha(t)=-\alpha_{0}\sin\nu t,$ $\alpha_{0}=eA_{0}$v$_{x}/m_{e}.$ Replacing
the modified Hamiltonian in eq. (\ref{Schro}), we obtain:
\begin{align}
\overset{\cdot}{A}(t) &  =-iA\omega_{a}+i\Omega_{R}B\cos\nu t+iA\alpha_{0}%
\sin\nu t,\\
\overset{\cdot}{B}(t) &  =-iB\omega_{b}+i\Omega_{R}A\cos\nu t+iB\alpha_{0}%
\sin\nu t.
\end{align}
where it was used $P_{ab}=P_{ab}^{\ast}.$ In terms of the slowing varying
amplitudes:
\begin{align}
\overset{\cdot}{a}(t) &  =i\alpha_{0}a(t)\sin\nu t+i(\Omega_{R}%
/2)b(t)e^{i(\omega-\nu)t},\label{a2}\\
\overset{\cdot}{b}(t) &  =i\alpha_{0}b(t)\sin\nu t+i(\Omega_{R}%
/2)a(t)e^{-i(\omega-\nu)t},\label{b2}%
\end{align}
At resonance $\left(  \omega=\nu\right)  $, and with the initial condition
$a(0)=1,$ such differential equations can be exactly solved, yielding
\begin{align}
a(t) &  =e^{-i(\alpha_{0}\cos\nu t)/\nu}e^{i\alpha_{0}/\nu}\cos\left(
\frac{\Omega_{R}t}{2}\right)  ,\label{sol2a}\\
b(t) &  =e^{-i(\alpha_{0}\cos\nu t)/\nu}e^{i\alpha_{0}/\nu}\sin\left(
\frac{\Omega_{R}t}{2}\right)  .\label{sol2b}%
\end{align}
These coefficients are different from the former ones (Eqs. (\ref{sol1a}) and
(\ref{sol1b})), but the amplitude probabilities $|A(t)|^{2},|B(t)|^{2}$ and
the population inversion (PI), $W(t)=\cos\Omega_{R}t,$ are not altered. This
shows that the Lorentz-violating term $\mathbf{A}\cdot\mathbf{v}$\ does not
modify the Rabi oscillation and the physics of the two-level system, except
for phase effects.

We should now investigate the effect of term $\mathbf{p}\cdot\mathbf{v}$ on
the two-level system. In this case, the Hamiltonian is: $H=H_{0}+H_{int}%
+H_{2},$ where $H_{2}=-(\mathbf{p}\cdot\mathbf{v})/m_{e}.$ For an x-axis
polarized electric field, the electron momentum should also be aligned along
the x-axis. In this case: $\ H_{2}=-$v$_{x}p_{x}/m_{e}.$ Using the relation
$\overset{\cdot}{x}=$ $-i[x,H_{0}],$ it results: $H_{2}=i$v$_{x}[x,H_{0}].$
Representing such operator in the $\{|a\rangle,|b\rangle\}$\ basis, we
have$\ H_{2}=-i\beta_{0}[\omega|a\rangle\langle b|-\omega|b\rangle\langle
a|],$ with $\beta_{0}=($v$_{x}p_{ab}),$ $p_{ab}=\langle a|x|b\rangle.$
Replacing all that into the Schr\"{o}dinger equation, we get the following
system of coupled differential equations:%
\begin{align}
\overset{\cdot}{A}(t) &  =-i\omega_{a}A+iP_{ab}E_{0}B\cos\nu t-\beta_{0}\omega
B,\label{a3A}\\
\overset{\cdot}{B}(t) &  =-i\omega_{b}B+iP_{ab}E_{0}A\cos\nu t+\beta_{0}\omega
A.\label{b3A}%
\end{align}
Writing such equations for the slowing varying amplitudes (in the rotating
wave approximation), we attain:
\begin{align}
\overset{\cdot}{a}(t) &  =i\frac{P_{ab}E_{0}}{2}b(t)e^{i(\omega-\nu)t}%
-\beta_{0}\omega b(t)e^{i\omega t},\label{a3B}\\
\overset{\cdot}{b}(t) &  =i\frac{P_{ab}E_{0}}{2}a(t)e^{-i(\omega-\nu)t}%
+\beta_{0}\omega a(t)e^{-i\omega t},\label{b3B}%
\end{align}

An interesting preliminary analysis consists in analyzing the behavior of this
system under the action\ only of the background, in a situation where the
external electromagnetic field is null. In this case, it is possible to show
that the Lorentz-violating background is able to induce Rabi oscillations,
once the system has a non null electric dipole moment. In the absence of the
external field, the system of Eqs. (\ref{a3B}), (\ref{b3B}) takes the form
$\overset{\cdot}{a}(t)=-\beta_{0}\omega b(t)e^{i\omega t},$ $\overset{\cdot
}{b}(t)=\beta_{0}\omega a(t)e^{-i\omega t},$ which implies the
solution:\textbf{ }%
\begin{equation}
a(t)=\frac{1}{2\sqrt{1+4\beta_{0}^{2}}}\left[  k_{-}e^{\frac{i}{2}k_{-}\omega
t}+k_{+}e^{-\frac{i}{2}k_{+}\omega t}\right]  ,b(t)=\frac{-i\beta_{0}}%
{\sqrt{1+4\beta_{0}^{2}}}\left[  e^{\frac{i}{2}k_{-}\omega t}-e^{-\frac{i}%
{2}k_{+}\omega t}\right]  ,
\end{equation}
with $k_{\pm}=(\sqrt{1+4\beta_{0}^{2}}\pm1).$ The corresponding population
inversion is%
\begin{equation}
W(t)=\frac{1}{\left(  1+4\beta_{0}^{2}\right)  }\left[  1+4\beta_{0}^{2}%
\cos(\sqrt{1+4\beta_{0}^{2}}\omega t)\right]  . \label{WV}%
\end{equation}
This result shows that the Lorentz-violating background, by itself, is able to
induce state oscillations with a fixed frequency $\varpi$, that approximates
to \textbf{ }$\omega=(\omega_{a}-\omega_{b}),$ once the background is supposed
to be of small magnitude $\left(  \beta_{0}^{2}<<1\right)  $. \textbf{ }As the
corrections induced are proportional to the factor $4$v$_{x}{}^{2}p_{ab}^{2},$
it may then be used to impose a bound on the background magnitude. In fact,
considering that effects on the population inversion larger than $10^{-10}$
might be experimentally detectable, we shall have $4$v$_{x}{}^{2}p_{ab}%
^{2}<10^{-10}.$ Taking $p_{ab}=\langle a|x|b\rangle\simeq1(eV)^{-1},$ it
yields: v$_{x}{}<5\cdot10^{-6}eV.$ This is not a tight upper bound, but may be
taken as a preliminary result. A more stringent bound can be attained
analyzing the system behavior in the presence of electric field $\left(
E_{0}\neq0\right)  $.

In the presence of \ the external field $E_{0}$, Eqs. (\ref{a3B},\ref{b3B}) do
not possess analytical solution even at resonance $\left(  \omega=\nu\right)
$. A numerical approach is then employed to provide a graphical solution for
the PIF of the system. To do it, it is necessary to establish a set of
numerical values compatible with the physics of a typical two-level system.
Here, we are working at the natural units ($c=1,\hbar=1),$ where the relevant
parameters present the following mass dimension: $[\Omega_{R}]=[\omega
]=[v^{\mu}]=1,[E_{0}]=2,[P_{ab}]=[p_{ab}]=-1.$

As a starting step, we search for the effects of a small background,
$|$\textbf{v}$|=10^{-6}-10^{-8}eV,$ on the two-level system. We take values
for the electric field corresponding to a typical magnetic field $\left(
B_{0}=10^{-4}-10^{-3}T\Rightarrow E_{0}=3\times10^{3}-3\times10^{4}%
Volt/m\right)  .$ In natural units $[1$ $volt/m=2.3\times10^{-6}(eV)^{2}],$ so
it results: $E_{0}\simeq0.7-7(eV)^{2}.$ For the wave frequency, we take a
typical electromagnetic value: $\nu\sim10^{16}Hz$ (ultraviolet limit), which
in natural units is equivalent to $\nu=6.6eV$ $\ ($since $1s^{-1}%
=6.6\times10^{-16}eV).$

\begin{figure}
[ptb]
\begin{center}
\includegraphics[
height=1.8421in,
width=5.2382in
]%
{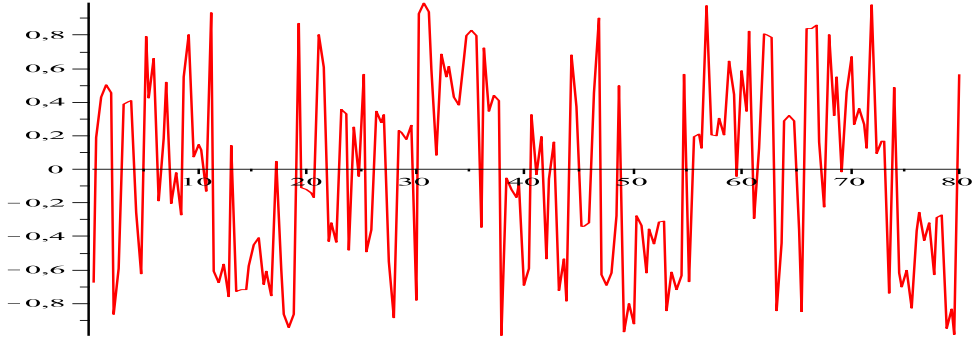}%
\caption{Population Inversion versus time plot for $P_{ab}=3(eV)^{-1}%
,\nu=3eV,$\textbf{v} $=10^{-7}eV,$ $E_{0}=2(eV)^{2}.$}%
\label{Fig1}%
\end{center}
\end{figure}

\begin{figure}
[ptb]
\begin{center}
\includegraphics[
height=1.8524in,
width=5.2382in
]%
{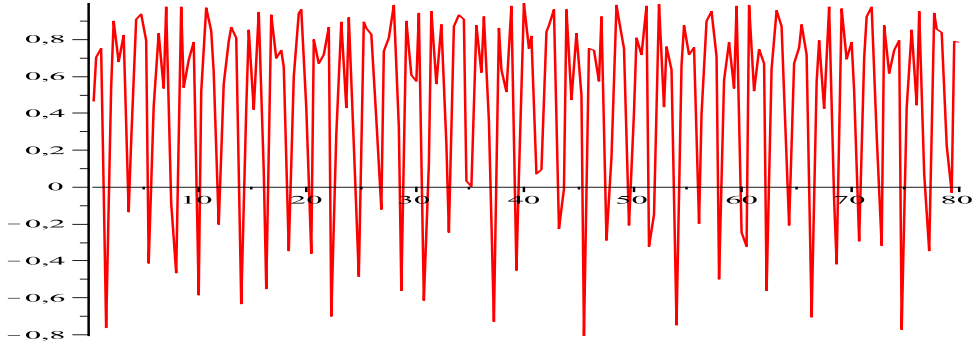}%
\caption{Population Inversion versus time plot for $P_{ab}=1(eV)^{-1}%
,\nu=3eV,$\textbf{v} $=10^{-7}eV,$ $E_{0}=2(eV)^{2}.$}%
\label{Fig2}%
\end{center}
\end{figure}

In Fig. [\ref{Fig1}], the effect of the background on the PIF is shown in
detail. It induces alterations (peak deformations) on the perfect harmonic
sinusoidal pattern, given by eq. (\ref{W2}). Such modifications appear at the
form of peak reversions and nonhomogeneities in the PIF. These alterations are
present along with all sinusoidal oscillation pattern. In the absence of an
analytical solution, it is necessary to scan the relevant parameters to gain
some feeling about the role played by each one. Naturally, Lorentz-violating
effects on the PIF increase with the background magnitude. Keeping $E_{0}%
,$v$,$ and the external frequency $\left(  \nu\right)  $ constant, several
numerical simulations revealed that the LV effects tend to diminish with
$P_{ab},$ as observed in Fig. [\ref{Fig2}]. The variation of the electric
field seems to have a cyclic effect on $W.$ Initially, while $E_{0}$ magnitude
is reduced, increasing LV effects are implied. Reducing even more $E_{0},$
such effects diminish, so that in the limit $E_{0}\rightarrow0,$ the
background influence becomes tiny, recovering the weak oscillations described
by eq. (\ref{WV}). Concerning the oscillation frequency, it continues to be
sensitive to the value of $P_{ab}E_{0}$ (the larger this product, the larger
the frequency), but such frequency is also affected by the background
magnitude, so that $\Omega_{R}=E_{0}P_{ab}$ does not hold anymore. Now, it
increases with the background, being larger than $E_{0}P_{ab}$. Only when the
background tends to vanish, the oscillation frequency recovers the usual value
$\left(  E_{0}P_{ab}\right)  .$%
%

Moreover, for some specific parameter values, the background induces a clear
modulation on the PIF, which takes place only for some values of the product
$P_{ab}E_{0}.$ Such a modulation is obviously associated with a kind of
partial inversion frustration (when the inversion is not fully accomplished)
at some stages. In fact, the graph of Fig. [\ref{Fig3}] shows a pattern of
modulation very similar to the one of a beat (superposition of close
frequencies). The behavior of Fig. [\ref{Fig3}] occurs for some specific
combinations of $P_{ab}$ and $E_{0}$ which yield $P_{ab}E_{0}=3,$ for
$\nu=3eV,$\textbf{v} $=10^{-8}eV.$ It was also reported for $\left(
P_{ab}=3/2,E_{0}=2\right)  $ and $\left(  P_{ab}=3/4,E_{0}=4\right)  ,$
keeping the values of $\nu$ and v unchanged. As already commented, a $P_{ab}$
reduction implies an attenuation on the LV effects, except when the resulting
value of $P_{ab}E_{0}$ brings about a beat oscillation pattern.\textbf{%
\begin{figure}
[ptb]
\begin{center}
\includegraphics[
height=2.092in,
width=5.2382in
]%
{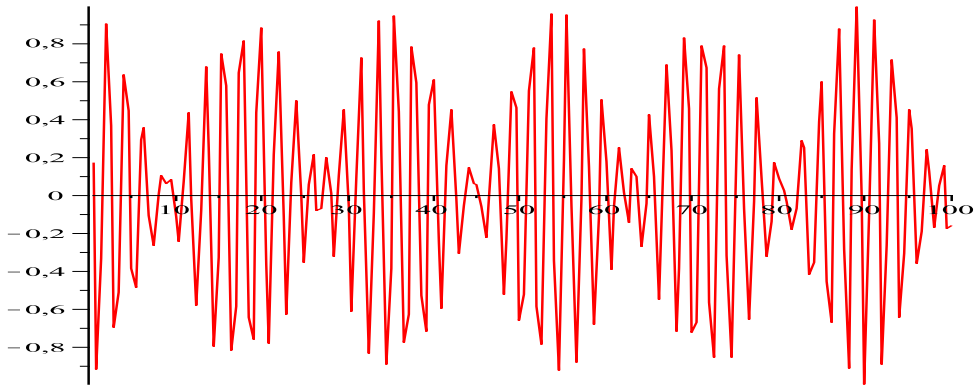}%
\caption{Population Inversion versus time plot for $P_{ab}=1(eV)^{-1}%
,\nu=3eV,$\textbf{v} $=10^{-8}eV,$ $E_{0}=3(eV)^{2}$}%
\label{Fig3}%
\end{center}
\end{figure}
}

Finally, an important point also to be analyzed refers to the minimum
background magnitude for which modifications on the PIF are still present.
Numerical simulations for $\mathbf{v}_{x}$ $=10^{-10}eV$ do not reveal any
effect on the usual sinusoidal Rabi oscillations for the following parameter
ranges: $0<P_{ab}<10\left(  eV\right)  ^{-1}$, $0<E_{0}<10\left(  eV\right)
^{2},0.001<\nu<6\left(  eV\right)  .$ Parameter values outside these ranges
could also be considered, but violate the dipolar approximation and are not
suitable to simulate the physics of a two-level system (except for frequencies
smaller than $0.001eV$). Once the Lorentz-violating modifications alluded to
here are not observed, we should conclude that the background magnitude can
not be larger than $10^{-10}eV$ $\left(  \mathbf{v}_{x}\leq10^{-10}eV\right)
$.

\subsection{Lorentz-violating effects due to the axial-vector coupling}

We now examine the effects that the coupling $b_{\mu}\overline{\psi}\gamma
_{5}\gamma^{\mu}\psi$ amounts to the two-level system. Initially, we choose a
purely spacelike background, $b^{\mu}=(0,\mathbf{b),}$ for which there
corresponds the nonrelativistic term $\mathbf{\sigma}\cdot\mathbf{b}$. In this
case, we shall adopt a four-state basis: $\left\{  |a+\rangle,|a-\rangle
,|b+\rangle,|b-\rangle\right\}  ,$ where each element is a tensor product of
an eigenstate of energy $(|a\rangle$ or $|b\rangle)$ with an eigenstate of
spin $(|+\rangle$ or $|-\rangle).$ In this case, the general state vector is%

\begin{equation}
|\psi(t)\rangle=A_{1}(t)|a_{+}\rangle+A_{2}(t)|a_{-}\rangle+B_{1}%
(t)|b+\rangle+B_{2}(t)|b-\rangle,
\end{equation}
which formally describes a four-level system. The Hamiltonian is
$H=H_{0}+H_{int}+H_{3},$ with $H_{3}=\mathbf{\sigma}\cdot\mathbf{b}.$ In this
basis, we write:
\begin{align}
H_{0}  &  =\hbar\omega_{a}(|a+\rangle\langle a+|+|a-\rangle\langle
a-|)+\hbar\omega_{b}(|b+\rangle\langle b+|+|b-\rangle\langle b-|),\label{H0}\\
H_{int}  &  =-(P_{ab++}|a+\rangle\langle b+|+P_{ab--}|a-\rangle\langle
b-|+P_{ab++}^{\ast}|b+\rangle\langle a+|+P_{ab--}^{\ast}|b-\rangle\langle
a-|)E(t),\label{Hint}\\
H_{3}  &  =\{b_{x}[|a-\rangle\langle a+|+|a+\rangle\langle a-|+|b-\rangle
\langle b+|+|b+\rangle\langle b-|]+b_{z}[|a+\rangle\langle a+|-|a-\rangle
\langle a-|\nonumber\\
&  +|b+\rangle\langle b+|-|b-\rangle\langle b-|]-ib_{y}[|a+\rangle\langle
a-|-|a-\rangle\langle a+|+|b+\rangle\langle b-|-|b-\rangle\langle b+|]\},
\label{H3}%
\end{align}
where it was taken: $H_{3}=b_{x}\mathbf{\sigma}_{x}+b_{y}\mathbf{\sigma}%
_{y}+b_{z}\mathbf{\sigma}_{z},$ $P_{ab++}=e\langle a+|x|b+\rangle
,P_{ab--}=e\langle a-|x|b-\rangle,$ and $\langle a+|x|b-\rangle=\langle
a-|x|b+\rangle=\langle b-|x|a+\rangle=\langle b+|x|a-\rangle=0,$ once the
operator $x$ does not act on spin states. The spin operators acts as
$\mathbf{\sigma}_{x}|\pm\rangle$ $=|\mp\rangle,$ $\mathbf{\sigma}_{z}%
|\pm\rangle$ $=\pm|\pm\rangle.$ Replacing all that in the Schr\"{o}dinger
equation, we obtain four coupled differential equations for the time-dependent coefficients:%

\begin{align}
\overset{\cdot}{A}_{1}  &  =-i\left[  \omega_{a}A_{1}-E(t)P_{ab++}B_{1}%
+b_{x}A_{2}+b_{z}A_{1}-ib_{y}A_{2}\right]  ,\label{diffcb1}\\
\overset{\cdot}{A}_{_{2}}  &  =-i\left[  \omega_{a}A_{2}-E(t)P_{ab--}%
B_{2}+b_{x}A_{1}-b_{z}A_{2}+ib_{y}A_{1}\right]  ,\\
\overset{\cdot}{B}_{_{1}}  &  =-i\left[  \omega_{b}B_{1}-E(t)P_{ab++}^{\ast
}A_{1}+b_{x}B_{2}+b_{z}B_{1}-ib_{y}B_{2}\right]  ,\\
\overset{\cdot}{B}_{_{2}}  &  =-i\left[  \omega_{b}B_{2}-E(t)P_{ab--}^{\ast
}A_{2}+b_{x}B_{1}-b_{z}B_{2}+ib_{y}B_{1}\right]  , \label{diffcb4}%
\end{align}

Such equations may be written in terms of the slow varying amplitudes
$(a_{1}(t)=A_{1}e^{i\omega_{a}t},a_{2}(t)=A_{2}e^{i\omega_{a}t},b_{1}%
(t)=B_{1}e^{i\omega_{b}t},b_{2}(t)=B_{2}e^{i\omega_{b}t})$:%
\begin{align}
\overset{\cdot}{a}_{1}(t)  &  =-i\left[  -E_{0}P_{ab}e^{i(\omega-\nu)t}%
b_{1}(t)+b_{x}a_{2}(t)+b_{z}a_{1}(t)-ib_{y}a_{2}\right]  ,\label{diffb1}\\
\overset{\cdot}{a}_{2}(t)  &  =-i\left[  -E_{0}P_{ab}e^{i(\omega-\nu)t}%
b_{2}(t)+b_{x}a_{1}(t)-b_{z}a_{2}(t)+ib_{y}a_{1}\right]  ,\\
\overset{\cdot}{b}_{1}(t)  &  =-i\left[  -E_{0}P_{ab}e^{-i(\omega-\nu)t}%
a_{1}(t)+b_{x}b_{2}(t)+b_{z}b_{1}(t)-ib_{y}b_{2}\right]  ,\\
\overset{\cdot}{b}_{2}(t)  &  =-i\left[  -E_{0}P_{ab}e^{-i(\omega-\nu)t}%
a_{2}(t)+b_{x}b_{1}(t)-b_{z}b_{2}(t)+ib_{y}b_{1}\right]  , \label{diffb4}%
\end{align}
where it was assumed that the dipolar transitions are real quantities and the
same between states of different spin polarizations, that is, $P_{ab++}%
=P_{ab--}=P_{ab}=P_{ab}^{\ast}.$ Moreover, the term $e^{i(\omega+\nu)t}$ was
neglected due to the RWA. Considering a general situation, it may occur both
spin\ state oscillations and energy\ state oscillations, so that it is
necessary to define two population inversion functions - a spin PIF $\left(
W_{S}\right)  $ and an energy PIF $\left(  W_{E}\right)  $:
\begin{align}
W_{S}  &  =|a_{1}(t)|^{2}+|b_{1}(t)|^{2}-|a_{2}(t)|^{2}-|b_{2}(t)|^{2}%
,\label{WS}\\
W_{E}  &  =|a_{1}(t)|^{2}+|a_{2}(t)|^{2}-|b_{1}(t)|^{2}-|b_{2}(t)|^{2}.
\label{WE}%
\end{align}
As a first insight, this model is to be regarded in the absence of the
Lorentz-violating background $(b_{x}=b_{y}=b_{z}=0)$ and in the presence of
the external field $E_{0}.$ In this case, the solution for the four coupled
equations has the form:
\begin{equation}
a_{1}(t)=\cos(E_{0}P_{ab}t),a_{2}(t)=0,b_{1}(t)=-i\sin(E_{0}P_{ab}%
t),b_{2}(t)=0,
\end{equation}
which implies the following PIFs:
\begin{equation}
W_{E}=|a_{1}(t)|^{2}-|b_{1}(t)|^{2}=\cos(2E_{0}P_{ab}t),\text{
\ \ \ \ \ \ \ \ }W_{S}=1. \label{Wb1}%
\end{equation}
These results indicate an energy eigenstate oscillation $\left(  a_{1}%
(t)\neq0,b_{1}(t)\neq0,a_{2}(t)=b_{2}(t)=0\right)  $, generated by the
external field, and total absence of spin oscillation (the field does not act
on the spin states), denoted by $W_{S}=1.$

On the other hand, we can search for the solution of this system in the
absence of external field $(E_{0}=0)$, for the background configuration
$\mathbf{b}=(b_{x},0,b_{z})$:%

\begin{equation}
a_{1}(t)=\cos(\sqrt{b_{x}^{2}+b_{z}^{2}}t)-i\frac{b_{z}}{\sqrt{b_{x}^{2}%
+b_{z}^{2}}}\sin(\sqrt{b_{x}^{2}+b_{z}^{2}}t),a_{2}(t)=-i\frac{b_{z}}%
{\sqrt{b_{x}^{2}+b_{z}^{2}}}\sin(\sqrt{b_{x}^{2}+b_{z}^{2}}t),b_{1}%
(t)=0,b_{2}(t)=0.
\end{equation}
Such relations obviously reflect an oscillation on the spin states, once the
system undergoes alternation between $|a+\rangle,|a-\rangle$ states, and the
inexistence of energy state oscillation. The obtained PIFs,
\begin{equation}
W_{S}=\cos^{2}(\sqrt{b_{x}^{2}+b_{z}^{2}}t)+\left[  \frac{b_{z}^{2}-b_{x}^{2}%
}{b_{z}^{2}+b_{x}^{2}}\right]  \sin^{2}(\sqrt{b_{x}^{2}+b_{z}^{2}}t),\text{
}W_{E}=1,
\end{equation}
confirm that this is really the case. For ($b_{x}=0,b_{z}\neq0),$ we have
$W_{S}=1$ (no spin oscillation). For ($b_{x}\neq0,b_{z}=0),$ we have
$W_{S}=\cos(2b_{x}t).$ This latter outcome shows that the background may
itself induce spin oscillation (even in the absence of external field), which
may be used to establish an upper bound on $b_{x}.$

We should now solve the system of equations (\ref{diffb1}-\ref{diffb4}) in the
presence both of the Lorentz-violating background and external field\textbf{
}$E_{0}$. As a starting situation, we regard the background as $\mathbf{b}%
=(0,0,b_{z}),$ which provides:%
\begin{equation}
a_{1}(t)=\frac{1}{2}e^{-i(b_{z}-E_{0}P_{ab})t}+\frac{1}{2}e^{-i(b_{z}%
+E_{0}P_{ab})t},a_{2}(t)=0,b_{1}(t)=-\frac{1}{2}e^{-i(b_{z}-E_{0}P_{ab}%
)t}+\frac{1}{2}e^{-i(b_{z}+E_{0}P_{ab})t},b_{2}(t)=0.
\end{equation}
It is easy to note that the population inversion of the energy states is still
equal to $\cos(2E_{0}P_{ab}t)$, showing that the term $b_{z}\mathbf{\sigma
}_{z}$ does not modify the population inversion of this system. Furthermore,
it does not yield spin state oscillation as well $\left(  W_{S}=1\right)  $,
which is consistent with the fact that the operator $\mathbf{\sigma}_{z}$ does
not flip the spin.

Once the influence of the $b_{z}$-component is understood, the role played by
a more general background, $b=(b_{x},b_{y},0),$ should be now analyzed. In
this case, the spin inversion is an expected result. The system of Eqs.
(\ref{diffb1}-\ref{diffb4}) also exhibits an exact solution, namely,%

\begin{align}
a_{1}(t)  &  =\frac{1}{2}[\cos Mt+\cos Nt],\text{ }a_{2}(t)=-\frac
{i(b_{x}+ib_{y})}{2b}[\sin Nt-\sin Mt],\nonumber\\
b_{1}(t)  &  =\frac{i}{2}[\sin Mt+\sin Nt],\text{ }b_{2}(t)=-\frac
{i(b_{x}+ib_{y})}{2b}[\cos Nt-\cos Mt].
\end{align}
where\textbf{ }$M=(E_{0}P_{ab}-b),N=(E_{0}P_{ab}+b),b=\sqrt{b_{x}^{2}%
+b_{y}^{2}}$. Carrying out the spin and energy PIF, defined in Eqs.(\ref{WS}%
,\ref{WE}), we get%

\begin{align}
W_{E}  &  =\cos(2E_{0}P_{ab}t),\label{WE2}\\
W_{S}  &  =\cos(2bt). \label{WS2}%
\end{align}
It is seen that the terms $b_{x}\mathbf{\sigma}_{x},b_{y}\mathbf{\sigma}_{y}$
do not modify $W_{E},$ that is governed only by the external field, whereas
the external field does not disturb the spin oscillation, which depends only
on the background. The spin population inversion has period $\pi/b,$ so that
the smaller the parameter $b,$ the larger is such period. For a tiny $b$ the
period may be so large that the inversion could become unobservable. Here, we
take a case where such spin oscillation is undetectable (period bigger than
$10^{4}s)$ to establish an upper bound for the background: $b<10^{-19}eV.$
This upper bound holds equally as $b_{x}<10^{-19}eV$ or $b_{y}<10^{-19}eV$ for
the cases $b_{y}=0$ or $b_{x}=0$, respectively. It is clear that the
background $b=(0,b_{y},0)$ induces effects totally similar to the ones of
$b=(b_{x},0,0)$, thus requiring no special attention.

Another situation of possible interest is $\mathbf{b}=(b_{x},0,b_{z}),$ for
which the energy oscillation might be correlated with the spin oscillation. In
this case, the system of Eqs. (\ref{diffb1}-\ref{diffb4}) provides the
following solution:\textbf{ }%
\begin{align}
a_{1}(t)  &  =\frac{1}{2bMN}[ib_{z}E_{0}P_{ab}(M\sin Nt-N\sin Mt)-ib_{z}%
b(N\sin Mt+M\sin Nt)+MNb(\cos Mt+\cos Nt)],\\
b_{1}(t)  &  =-\frac{1}{2bMN}[-ib^{2}(M\sin Nt-N\sin Mt)+iE_{0}P_{ab}b(N\sin
Mt+M\sin Nt)+b_{z}MN(\cos Nt-\cos Mt)],\text{ }\\
\text{ }a_{2}(t)  &  =-\frac{ib_{x}}{2b}[\sin Nt-\sin Mt],\text{ \ \ }%
b_{2}(t)=\frac{b_{x}}{2b}[\cos Mt-\cos Nt],
\end{align}
where $b=\sqrt{b_{x}^{2}+b_{z}^{2}}.$ These outcomes allow one to write
lengthy expressions for $W_{E}$ and $W_{S}$ which reveal the influence of
$b_{x}$ and $b_{z}$ on the dynamics of the system. PIF graphs for several
values of $b_{x}$ and $b_{z}$ show that these coefficients do not have much
effect on the energy inversion, which remains almost invariant while $b_{x}$
and $b_{z}$ take on different values. This is not the case for the spin
inversion, however. Indeed, $W_{S}$ is sensitive to variations of the ratio
$b_{x}/b_{z},$ being suppressed for small values of $b_{x}/b_{z}$ $\left(
b_{x}/b_{z}<<1\right)  $, while increasing with it, becoming total $\left(
-1\leq W_{S}\leq1\right)  $ for the case $b_{x}/b_{z}>>1$. On the other hand,
it is also observed that the frequency of the spin oscillation increases with
the background modulus, $b.$ Given that the coefficient $b_{y}$ plays a role
similar to the one of $b_{x},$ we should conclude that the case $b=(0,b_{y}%
,b_{z})$ presents the same general behavior of this previous case.

\begin{figure}
[ptb]
\begin{center}
\includegraphics[
height=2.0816in,
width=5.2382in
]%
{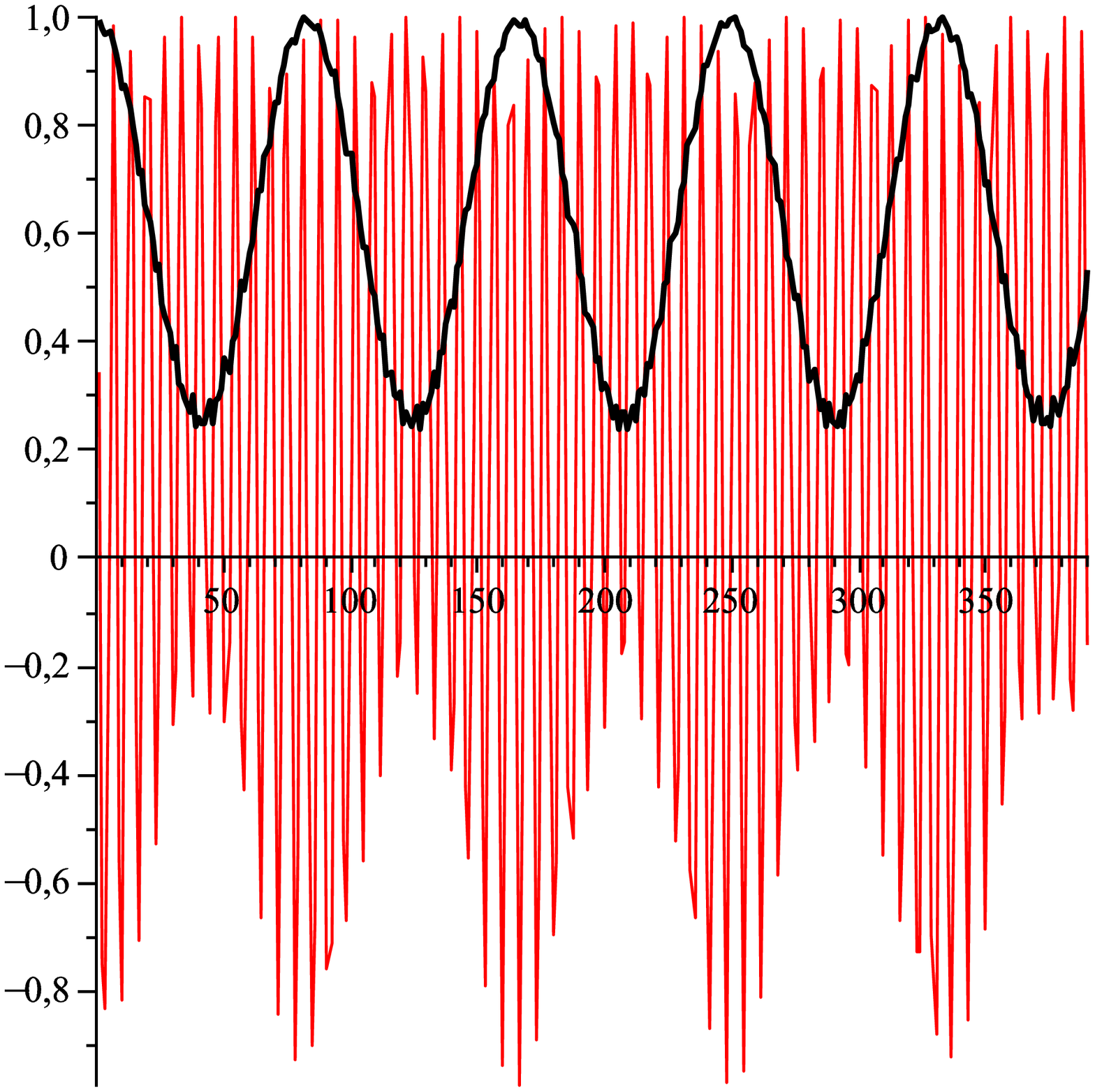}%
\caption{Simultaneous plot of spin PIF (black thick line) and energy PIF (thin
line) for the following parameter values: $P_{ab}=0.47\left(  eV\right)
^{-1},b_{0}=0.10eV,E_{0}=1.00(eV)^{2},\nu=1.00eV.$}%
\label{Fig4}%
\end{center}
\end{figure}

\begin{figure}
[ptb]
\begin{center}
\includegraphics[
height=1.9355in,
width=5.2382in
]%
{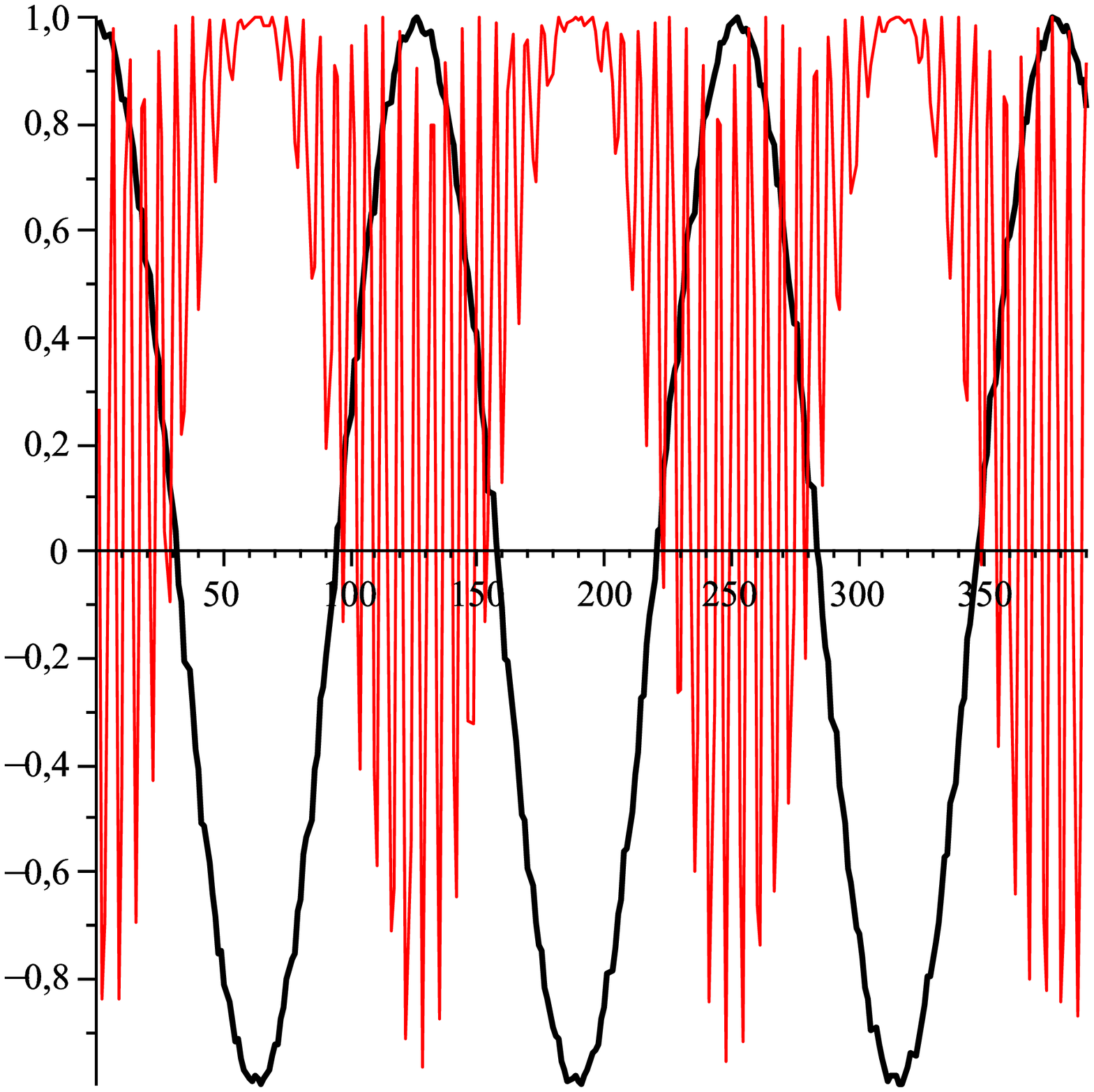}%
\caption{Simultaneous plot of spin PIF (black thick line) and energy PIF (thin
line) for the following parameter values: $P_{ab}=0.5\left(  eV\right)
^{-1},b_{0}=0.1eV,E_{0}=1.0(eV)^{2},\nu=1.0eV.$}%
\label{Fig5}%
\end{center}
\end{figure}

\begin{figure}
[ptb]
\begin{center}
\includegraphics[
height=2.0505in,
width=5.2382in
]%
{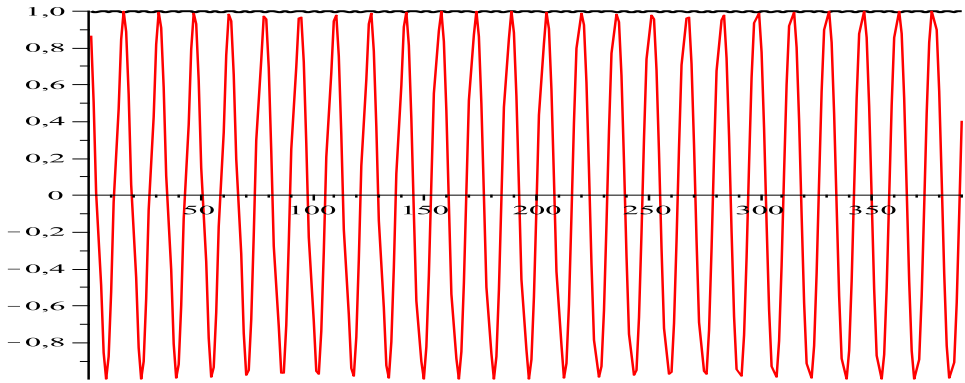}%
\caption{Simultaneous plot of spin PIF (black thick line) and energy PIF (thin
line) for the following parameter values: $P_{ab}=0.2\left(  eV\right)
^{-1},b_{0}=0.1eV,E_{0}=1.0(eV)^{2},\nu=1.0eV.$}%
\label{Fig6}%
\end{center}
\end{figure}

As a final investigation, we consider the case of a purely timelike
background, $b^{\mu}=(b_{0},\mathbf{0),}$ in which the corresponding
nonrelativistic term is $\mathbf{\sigma}\cdot(\mathbf{p}-e\mathbf{A})$.
Proceeding as earlier, the Hamiltonian is: $H=H_{0}+H_{int}+H_{4},$ with
$H_{4}=-(b_{0}/m_{e})\mathbf{\sigma}\cdot(\mathbf{p}-e\mathbf{A}).$ In the
basis $\left\{  |a+\rangle,|a-\rangle,|b+\rangle,|b-\rangle\right\}  $, the
interaction $H_{4}$ takes the form:
\begin{align}
H_{4}  &  =-ib_{0}\omega(P_{ab}^{\ast}|b-\rangle\langle a+|+P_{ab}^{\ast
}|b+\rangle\langle a-|-P_{ab}|a-\rangle\langle b+|-P_{ab}|a+\rangle\langle
b-|)\nonumber\\
&  +\gamma_{0}\sin(\nu t)[(|a+\rangle\langle a-|+|a-\rangle\langle
a+|+|b+\rangle\langle b-|+|b-\rangle\langle b+|],
\end{align}
where $\gamma_{0}=eb_{0}E_{0}/(m_{e}\nu),$ and $H_{0},H_{int}$ are already
written in Eqs. (\ref{H0}, \ref{Hint}). Replacing the full Hamiltonian in the
Schr\"{o}dinger equation, the following system of coupled equations is
obtained:
\begin{align}
\overset{\cdot}{a}_{1}(t)  &  =iP_{ab}\left[  E_{0}\cos(\nu t)e^{i\omega
t}b_{1}(t)-ib_{0}\omega e^{i\omega t}b_{2}(t)-\gamma_{0}(\sin\nu
t)a_{2}(t)\right]  ,\label{ab1}\\
\overset{\cdot}{a}_{2}(t)  &  =iP_{ab}\left[  E_{0}\cos(\nu t)e^{i\omega
t}b_{2}(t)-ib_{0}\omega e^{i\omega t}b_{1}(t)-\gamma_{0}(\sin\nu
t)a_{1}(t)\right]  ,\label{ab2}\\
\overset{\cdot}{b}_{1}(t)  &  =iP_{ab}\left[  E_{0}\cos(\nu t)e^{-i\omega
t}a_{1}(t)+ib_{0}\omega e^{-i\omega t}a_{2}(t)-\gamma_{0}(\sin\nu
t)b_{2}(t)\right]  ,\label{ab3}\\
\overset{\cdot}{b}_{2}(t)  &  =iP_{ab}\left[  E_{0}\cos(\nu t)e^{-i\omega
t}a_{2}(t)+ib_{0}\omega e^{-i\omega t}a_{1}(t)-\gamma_{0}(\sin\nu
t)b_{1}(t)\right]  , \label{ab4}%
\end{align}
Such a system does not provide an analytical solution, so that a numerical
approach must be employed. It is important to point out that the constant
$\gamma_{0}$ is much smaller than $E_{0},P_{ab},$ in such a way that the term
linear in $\gamma_{0}$ turns out negligible in comparison with the others.
Following the example of the last section, we solve the system (\ref{ab1}%
-\ref{ab4}) numerically. The graph of Fig. [\ref{Fig4}] depicts the behavior
of the PIF for the energy and spin states (thicker black line). It shows an
appreciable modification on the eigenenergy inversion induced by the
background - a partial frustration at some moments, whereas the spin inversion
is always partially frustrated (the system remains predominantly in the state
$|+\rangle$). This effect may\ be amplified if we take $P_{ab}=0.5\left(
eV\right)  ^{-1},$ as properly shown in Fig. [\ref{Fig5}], where the spin
oscillation becomes total and the energy inversion is entirely frustrated at
some moments. \ Numerical investigations have shown that this scenario takes
place only for specific values of the product $P_{ab}E_{0},$ (in this case,
$P_{ab}E_{0}=0.5\left(  eV\right)  ^{-1}).$ A quite similar behavior was
verified \ for $P_{ab}=0.25\left(  eV\right)  ^{-1},$ $b_{0}=0.10eV,$
$E_{0}=2.00(eV)^{2},$ $\nu=1.00eV.$ For values of $P_{ab}$ slightly different,
$P_{ab}=0.4\left(  eV\right)  ^{-1}$ or $P_{ab}=0.6\left(  eV\right)  ^{-1}$,
the spin oscillation is nearly annihilated, while the energy oscillation
becomes\ approximately total. This behavior is much similar to the one
described in Fig. [\ref{Fig6}]. The picture of Fig. [\ref{Fig5}] reveals an
interesting inversion pattern, observed for some background values. This graph
shows that the system is nearly collapsed at the state $|a-\rangle$ at a
moment, after undergoing oscillation between states $|a+\rangle,|b+\rangle$,
and turning back to the state $|a-\rangle$ in the sequel. This cycle of
alternations is repeated along with the time. By Fig. [\ref{Fig6}], one notes
that a $P_{ab}E_{0}$ reduction implies a lower frequency oscillation. This is
confirmed by analyzing several plots for different $E_{0},P_{ab}$ values.
\ Lastly, it was observed that the frequency of energy oscillation decreases
with $b_{0}$ magnitude as well.

Numerical simulations have shown that more appreciable LV effects are just
attainable for high values of the background, such as $b_{0}=0.1eV.$ For
smaller background magnitudes, $b_{0}\leq0.01eV$ (see Fig. [\ref{Fig6}]),
Lorentz-violating effects tend to vanish. Indeed, while the energy PIF tends
to assume the usual sinusoidal form, the spin PIF tends to collapse to 1
(absence of spin oscillation), in a behavior similar to the one of Fig.
[\ref{Fig6}].%

\section{Conclusion}

In this work, the effects of CPT- and Lorentz-violating terms (stemming from
the SME model) on a two-level system were investigated. In the first case
analyzed, it was reported that the background $v_{\mu}$ alters the Rabi
oscillations, implying partial population inversion frustration and some kinds
of modulation. Depending on the parameter values, the induced modulation might
be stronger, becoming similar to a beat. It also occurs that $v_{\mu}$ may
induce Rabi oscillations even in the absence of an external electromagnetic
field, but it would become a sensitive effect just for large background
magnitudes. Once the Lorentz-violating effects here foreseen are supposed to
be not observed, the minimum value below which the background\ yields no
modifications on the PIF was identified as an upper bound for this
Lorentz-violating coefficient ($\mathbf{v}_{x}$ $\leq10^{-10}eV).$ In order to
determine the effects induced by the background $b_{\mu},$ a four state basis
was adopted. It was then shown that it occurs two types of population
inversion, one referred to the energy states (determined by the external
electromagnetic field), another concerned with the spin states (implied by the
background). The supposed non observation of this spin oscillation (relying on
the validity of the dipole approximation) was used to impose an upper bound on
such a coefficient: $b_{x}<10^{-19}eV.$

A point that deserves some attention concerns to the validity of the rotating
wave approximation (RWA) in the cases where the Lorentz-violating perturbation
contains terms like sin$\nu t$ or cos$\nu t$, as in Eqs. (\ref{a2},
\ref{b2},\ref{ab1}-\ref{ab4}). \ In this case, the perturbation term
oscillates at a frequency $\left(  \nu\right)  $ equal to half the frequency
of the RWA-neglected term. The question is to know if the rapidly oscillating
term (with frequency equal to $2\nu$) may be dropped out while the
perturbation term (with frequency equal to $\nu$) is kept. To answer this
issue, some numerical calculations were performed out of the RWA, that is,
maintaining the rapidly oscillating terms together. The observed results do
not differ qualitatively and appreciably from the ones previously obtained,
which leads to the conclusion that RWA is still a good approximation.

Finally, the cases here studied in the semiclassical viewpoint can also be
considered in the context of a quantized electromagnetic field in a cavity. In
this case, for the $\mathbf{A}\cdot\mathbf{v}$ term the results depend on the
initial state of the electromagnetic field. For an initially excited atom in a
coherent state with large number of photons, the term induces only a phase on
the wavefunction coefficients. As it does not alter the probability
amplitudes, it corroborates the results obtained at Sec. IIIB (at
semi-classical level). We are now investigating the effects due to the
background for small number of photons in the cavity, such as corrections
induced on the population inversion and photon statistics. This work is under
development\textbf{ }\cite{Adalto}\textbf{.}

\begin{acknowledgments}
M.M. Ferreira Jr expresses his gratitude both to FAPEMA\ (Funda\c{c}\~{a}o de
Amparo \`{a} Pesquisa do Estado do Maranh\~{a}o) and to the CNPq (Conselho
Nacional de Desenvolvimento Cient\'{\i}fico e Tecnol\'{o}gico), A. R. Gomes is
grateful to CNPq, and R. C. C. Lopes acknowledges CAPES\ for financial
support. The authors thank B. Baseia for reading the manuscript and for
valuable suggestions.
\end{acknowledgments}

\end{document}